\documentclass[table]{article} 
\usepackage{iclr2025_conference,times}
\iclrfinalcopy

\usepackage{amsmath,amsfonts,bm}

\def\eqref#1{equation~\ref{#1}}

\def\1{\bm{1}}

\DeclareMathAlphabet{\mathsfit}{\encodingdefault}{\sfdefault}{m}{sl}
\SetMathAlphabet{\mathsfit}{bold}{\encodingdefault}{\sfdefault}{bx}{n}

\usepackage{hyperref}
\usepackage{url}
\usepackage{graphicx}
\usepackage{booktabs}
\usepackage{tabularx}
\usepackage{multirow}
\usepackage{amssymb}
\usepackage{cleveref}

\crefname{figure}{Figure}{Figures}
\crefname{table}{Table}{Tables}
\crefname{appendix}{Appendix}{}

\title{Tackling Few-Shot Segmentation \\in Remote Sensing \\via Inpainting Diffusion Model}

\author{Steve Andreas Immanuel, \;\; Woojin Cho, \;\; Junhyuk Heo, \;\; Darongsae Kwon \\
TelePIX \\
07330, Seoul, South Korea \\
\texttt{\{steve, woojin, hjh1037, darong.kwon\}@telepix.net} \\
}

\begin{document}

\maketitle

\begin{abstract}
Limited data is a common problem in remote sensing due to the high cost of obtaining annotated samples. In the few-shot segmentation task, models are typically trained on base classes with abundant annotations and later adapted to novel classes with limited examples. However, this often necessitates specialized model architectures or complex training strategies. Instead, we propose a simple approach that leverages diffusion models to generate diverse variations of novel-class objects within a given scene, conditioned by the limited examples of the novel classes. By framing the problem as an image inpainting task, we synthesize plausible instances of novel classes under various environments, effectively increasing the number of samples for the novel classes and mitigating overfitting. 
The generated samples are then assessed using a cosine similarity metric to ensure semantic consistency with the novel classes. 
Additionally, we employ Segment Anything Model (SAM) to segment the generated samples and obtain precise annotations. By using high-quality synthetic data, we can directly fine-tune off-the-shelf segmentation models. Experimental results demonstrate that our method significantly enhances segmentation performance in low-data regimes, highlighting its potential for real-world remote sensing applications. All the codes are publicly available at \url{https://github.com/SteveImmanuel/rs-paint}.
\end{abstract}

\section{Introduction}

Remote sensing is a task to capture images of Earth's surface from a distance, typically using satellites. These data can then be utilized for many applications, such as climate forecasting \citep{troccoli2010seasonal, palmer2014climate}, marine ecosystem monitoring \citep{kavanaugh2021satellite}, urban planning \citep{malarvizhi2016use}. Due to the very high dimensional nature of satellite data, one of the most crucial part to process these data is to locate and segment any area of interest within these images. There has been a plethora of works in developing algorithm for object detection \citep{guo2018geospatial,gong2022swin} and segmentation \citep{wu2019towards, bahl2019low, karimov2024deep} to automate this process, notably using deep neural network. 

However, like most neural networks, these models require extensive training data to achieve high performance. In the remote sensing domain, obtaining such datasets is particularly challenging. The images themselves are costly to acquire, often requiring access to specialized satellite systems or proprietary archives. Even when the data is available, privacy or security policies often restrict access to sensitive regions, e.g., military zones, further limiting usable training samples. Moreover, generating annotations is even more resource-intensive, as it requires domain expertise for accurate labeling, such as identifying environmental patterns, urban structures, or marine ecosystems. 
This combination of high costs and labor-intensive annotation processes significantly limits the availability of large-scale labeled datasets in the remote sensing domain. 

Several works \citep{liu2023learning,yang2023mianet,hajimiri2023strong} have focused on developing few-shot learning algorithms for semantic segmentation to enable models to perform well with limited data. While these methods offer promising results, they usually only work on specific settings or require complex training strategy. We argue that a more effective approach to address this issue is to circumvent the data scarcity problem altogether. To this end, we propose leveraging inpainting diffusion models to generate additional training samples including the annotations at minimal cost. By conditioning the diffusion model on the object of interest, we can generate a diverse set of synthetic images featuring that object across a variety of scenes. Subsequently, we employ SAM to automatically derive the corresponding semantic masks. Although other data augmentation techniques, such as Copy-Paste \citep{ghiasi2021simple}, have been explored, they have significant drawbacks. Specifically, while Copy-Paste can also increase the number of training samples, it often produces unrealistic images with noticeable artifacts along the object boundaries. When models are trained on such data, they may learn to rely on these artifacts as shortcuts for object detection or segmentation. Since these artifacts are absent in real-world images, the models trained in this manner tend to suffer from poor generalization performance.
\begin{figure}[t]
    \centering
    \includegraphics[width=\textwidth]{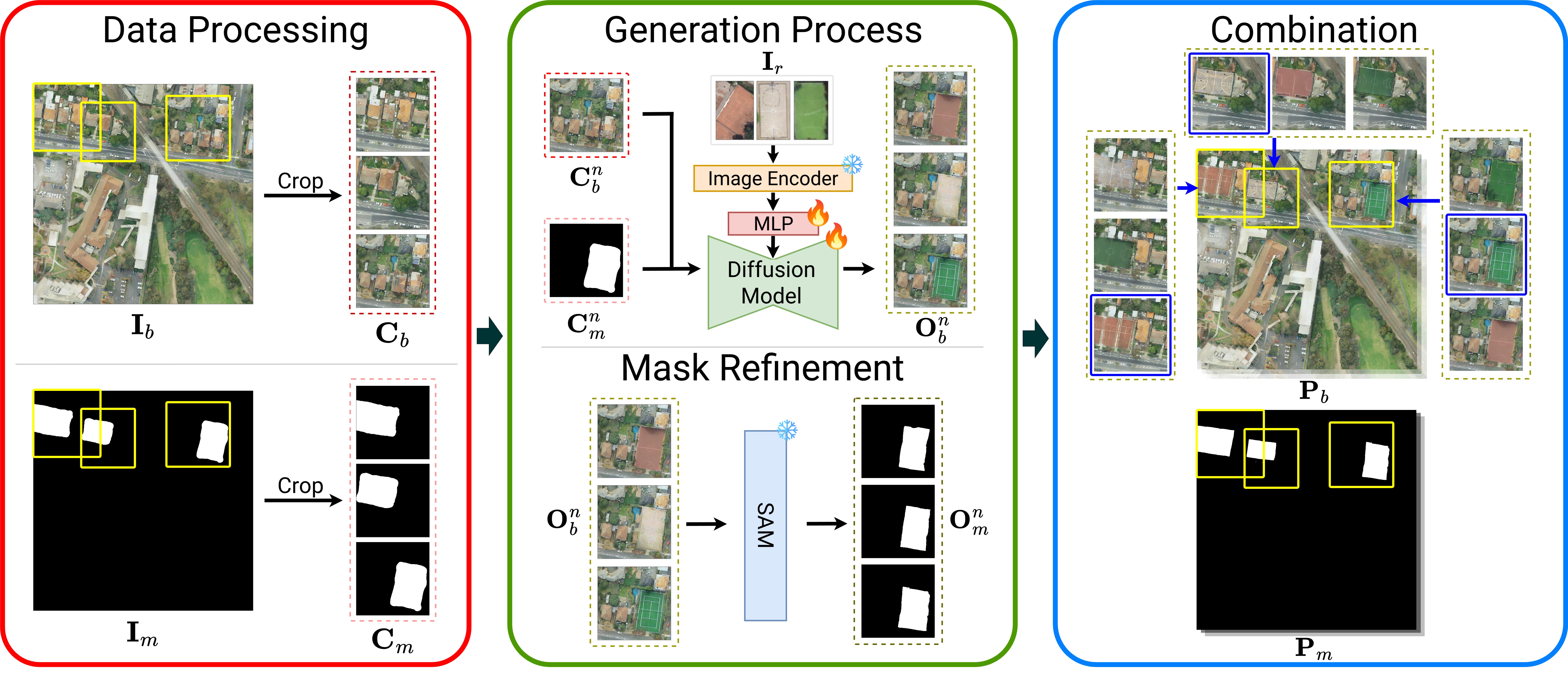}
    \caption{Overall pipeline of the proposed approach.  An inpainting diffusion model generates novel-class samples, and SAM refines the segmentation masks. The results are used for training samples to improve model performance on few-shot settings.}
    \label{fig:pipeline}
\end{figure}

\section{Preliminaries}

\paragraph{Few-shot segmentation.} In few-shot segmentation setting \citep{tian2022generalized}, there are base classes and novel classes. The model is trained on abundant samples from the base classes and then adapted to segment instances of novel classes. Each novel class has a support set with a few annotated examples and a query set consisting of images to be segmented.

\paragraph{Image inpainting.} Given a base image $\mathbf{I}_b \in \mathbb{R}^{H \times W \times 3}$, where $H$ and $W$ denote the height and width of the $\mathbf{I}_b$, and a binary mask $\mathbf{I}_m \in \{0, 1\}^{H \times W}$, with 0 and 1 incidating unmasked and masked areas, respectively, the goal of image inpainting task is to fill the masked area with pixels that are semantically and structurally coherent with the unmasked areas. In our approach, we leverage image inpainting model to generate the novel classes within the masked area. Therefore, we essentially increase the number of annotated samples by generating more variations of the novel classes.

\paragraph{Image-conditioned diffusion model.} Text-conditioned generative models, \textit{e.g.}, Stable Diffusion \citep{rombach2022high}, have demonstrated impressive performance in generating realistic images. However, text prompts can be ambiguous, especially in the remote sensing domain as objects often appear similar from a top-down perspective and are difficult to describe solely with text. To address this, \cite{yang2023paint} and \cite{song2023objectstitch} use a reference image $\mathbf{I}_r \in \mathbb{R}^{H' \times W' \times 3}$ instead of a text prompt to condition the generation process while maintaining the object identity in the reference image. Following \cite{yang2023paint}, $\mathbf{I}_r$ is processed through a frozen image encoder and compressed into a one-dimensional vector using multilayer perceptrons (MLP). This introduces information bottleneck which enforces the model to learn the semantic information of $\mathbf{I}_r$ without collapsing into trivial solution of copy-pasting $\mathbf{I}_r$ into $\mathbf{I}_b$. 

\section{Methodology}
Given the few examples in the support set of novel classes in segmentation dataset, we use Stable Diffusion to generate plausible variations of the novel classes in many different environments. The generated samples help train an off-the-shelf segmentation model, mitigating overfitting that occurs when training with only the support set. This approach eliminates the need for a specially designed few-shot segmentation model and simplifies training by avoiding the typical two-phase process \citep{tian2022generalized,liu2023learning,hajimiri2023strong}, where the model is first trained on base classes and then adapted to novel classes.

\subsection{Self-Supervised Training}
To train Stable Diffusion with an image prompt, we need to collect pairs of ($\mathbf{I}_b, \mathbf{I}_r, \mathbf{I}_m$) and the expected painted image $\mathbf{P}_b$. However, there are no publicly available datasets and it is infeasible to manually curate such dataset. Therefore, we leverage remote sensing object detection dataset for training in self-supervised manner. Given an image with a bounding box of an object in the image, we use the bounding box as the binary mask $\mathbf{I}_m$, the patch inside the bounding box as $\mathbf{I}_r$, and the original image as $\mathbf{P}_b$.

\subsection{Generation Process}
The overall generation pipeline is shown in \cref{fig:pipeline}. Given a base image $\mathbf{I}_b$ and its mask $\mathbf{I}_m$, there are $N$ plausible regions where an object can be generated. We first crop $\mathbf{I}_b$ into regions $\mathbf{C}_b=\{\mathbf{C}_b^n\}_{n=1}^{N}$ and $\mathbf{I}_m$ into their corresponding masks $\mathbf{C}_m=\{\mathbf{C}_m^n\}_{n=1}^{N}$. For each region $\mathbf{C}_b^n$, we independently generate $L$ different variations $\mathbf{O}_b^n = \{\mathbf{O}_{b}^{n,l}\}_{l=1}^L$ using $K$ different reference images $\{\mathbf{I}_{r}^{k}\}_{k=1}^K$. Due to the stochastic nature of Stable Diffusion, we can generate an arbitrary number of results even from a single reference image.

We find that the generated results are most realistic when the object mask covers approximately 15–30\% of the cropped region. To ensure high-quality synthesis, we compute the cosine similarity between the generated object and its corresponding reference image using a CLIP encoder. If the similarity score falls below a predetermined threshold, we regenerate the sample to balance semantic consistency with diversity. Each variation generated for different regions is then combined to form complete augmented images $\mathbf{P}_b$ and their masks $\mathbf{P}_m$. Given a single base image, the total number of unique variations that can be generated is $\sum_{k=1}^{N} \binom{N}{k}L^k$, where $\binom{N}{k}$ accounts for the selection of $k$ regions, and $L^k$ accounts for the variations generated per selected region. The full algorithm and derivation are provided in \cref{appendix:combine}.

\subsection{Mask Refinement}
The masks $\mathbf{C}_m^n$ act as guidance for the model on where to paint the novel class object. However, the generated object may be smaller than the mask, leading to inaccuracies. To obtain a more precise segmentation mask $\mathbf{O}_m^n$, we leverage SAM \citep{kirillov2023segment}, which has demonstrated strong zero-shot segmentation performance. While SAM does not inherently recognize object classes, this is not an issue in our case, as the generated object is conditioned on a reference image of a specific class, which effectively constraints the generated object class.

\section{Experiments}

\subsection{Dataset}
We use the few-shot set of the OpenEarthMap dataset \citep{broni_bediako_2024_11396874} and select four novel classes for benchmarking: \textit{boat}, \textit{agriculture land}, \textit{bridge}, and \textit{sportsfield}. This dataset was also used for the OpenEarthMap few-shot challenge \citep{broni2024generalized}. Each class has only 5 annotated samples. For our approach, we initialize the Stable Diffusion model from the pre-trained checkpoint provided by \cite{yang2023paint}. However, we replace the image encoder with pre-trained RemoteCLIP \citep{liu2024remoteclip} and fine-tune the entire model on the SAMRS dataset \citep{wang2024samrs} for 100 epochs. This model is then used to generate additional samples for the novel classes. Specifically, for each class, we generate approximately 1,000 new samples using the annotated samples as the image conditioning. The generated samples can be found in \cref{appendix:samples}. Additional dataset details can be found in \cref{appendix:dataset_details}.

\subsection{Performance Comparison}
As baselines, we choose YOLOv11 \citep{Jocher_Ultralytics_YOLO_2023}, SegFormer \citep{xie2021segformer}, and Mask2Former \citep{cheng2022masked} to represent different architectures. Each class is treated as a binary segmentation task, with a separate model trained for each class. We evaluate three training strategies: (i) Vanilla, using only the five annotated samples; (ii) Copy-Paste~\citep{ghiasi2021simple}, augmenting the five samples with copy-paste augmentation; and (iii) Ours, incorporating both the annotated and generated samples. More training details can be found in \cref{appendix:training_details}.

The results are presented in \cref{tab:comparison}. Models trained with only five annotated samples exhibit limited performance, often struggling with overfitting due to the scarcity of training data. Introducing the Copy-Paste augmentation improves performance in some cases, but it can also degrade results, as seen in the bridge class. This decline in performance is likely due to the inherent complexity of bridge objects, which need to connect two road segments in a structurally coherent manner. Simply copy-pasting bridge instances into new scenes does not guarantee a realistic connection between roads, potentially confusing the model and leading to suboptimal segmentation.

In contrast, our approach consistently delivers substantial improvements across all object classes and model architectures. By generating realistic novel-class samples through image inpainting, our method ensures that objects are naturally integrated into diverse environments, avoiding the pitfalls of naive augmentation techniques. This robustness underscores the flexibility and adaptability of our method, demonstrating its effectiveness regardless of the underlying segmentation model. Furthermore, we observe that models trained using our approach outperform can even outperform the challenge-winning submissions of the challenge. Importantly, our method achieves these results without relying on specialized architectures or complex training strategies, highlighting its practicality and ease of integration into existing workflows.

\begin{table}[t]
\caption{IoU comparison of various methods on different object classes. Results for challenge-winning methods are also included for reference. The \underline{underline} indicates best results in each group and the \textbf{boldface} indicates best results overall.}
\label{tab:comparison}
\centering
\footnotesize
\begin{tabularx}{\textwidth}{p{6cm}XXXX}
\specialrule{1pt}{2pt}{2pt}
\multirow{3.5}{*}{\textbf{Method}} & \multicolumn{4}{c}{\textbf{Class}}                                                                                 \\ 
\cmidrule(lr){2-5}
                                                          & \textbf{Boat}     & \textbf{Agriculture Land}  & \textbf{Bridge}   & \textbf{Sports Field}  \\
\specialrule{1pt}{2pt}{2pt}
YOLOv11 \citep{Jocher_Ultralytics_YOLO_2023}              & 5.40              & 9.37                       & 0.00              & 6.31              \\
+ Copy-Paste~\citep{ghiasi2021simple}                                              & 12.95             & 24.73                      & 0.00              & 24.85             \\
\rowcolor{lightgray} + Ours                               & \underline{47.39} & \textbf{\underline{46.35}} & \underline{45.96} & \underline{45.92} \\
\midrule
SegFormer \citep{xie2021segformer}                        & 6.00              & 26.49                     & 5.65              & 29.27             \\
+ Copy-Paste~\citep{ghiasi2021simple}                                              & 13.18             & 30.99                     & 4.94              & 29.08             \\
\rowcolor{lightgray} + Ours                               & \underline{45.02} & \underline{37.46}         & \underline{23.27} & \underline{45.10} \\
\midrule
Mask2Former \citep{cheng2022masked}                       & 10.63                      & 9.26              & 22.06                      & 20.86                      \\
+ Copy-Paste~\citep{ghiasi2021simple}                                              & 16.50                      & 36.19             & 14.63                      & 41.70                      \\
\rowcolor{lightgray} + Ours                               & \textbf{\underline{72.53}} & \underline{45.36} & \textbf{\underline{57.24}} & \textbf{\underline{66.36}} \\
\midrule
\multicolumn{5}{l}{\textit{Challenge Winners}}                                                                                                \\
\midrule
SegLand \citep{li2024generalized}                         & 10.76             & 8.22              & \underline{57.06} & \underline{55.87}     \\
ClassTrans \citep{wang2024class}                          & 0.00              & \underline{44.78} & 6.94              & 49.98                 \\
FoMA \citep{gao2024enrich}                                & \underline{58.64} & 1.44              & 17.01             & 40.59                 \\
P-SegGPT \citep{immanuel2024learnable}                    & 0.00              & 32.36             & 0.00              & 38.50                 \\
DKA \citep{tong2024dynamic}                               & 0.00              & 28.33             & 0.00              & 29.19                 \\
\specialrule{1pt}{2pt}{2pt}
\end{tabularx}
\end{table}

\section{Limitations}
\label{appendix:limitation}
\begin{figure}[ht]
    \centering
    \includegraphics[width=0.7\textwidth]{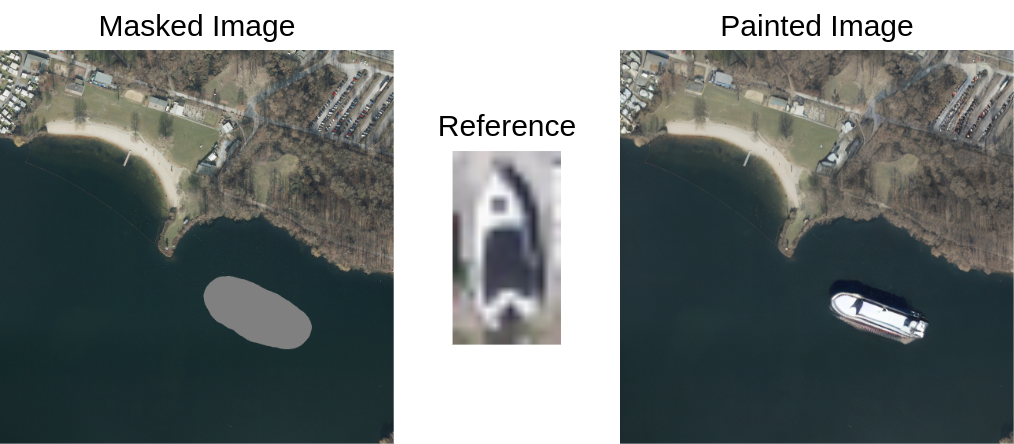}
    \caption{Failure case where the generated object is excessively large due to the large mask area.}
    \label{fig:limitation}
\end{figure}

Currently the mask $M$ needs to be manually created to ensure plausible location of the painted object. A promising direction for future work is to automate this process by leveraging a model to predict optimal object placement.

Additionally, the Stable Diffusion model does not account for the scale of objects relative to their surroundings. As shown in \cref{fig:limitation}, when a large mask area is used, the generated boat expands to fill most of the space. However, when compared to surrounding objects such as buildings, it is noticeably oversized. A straightforward idea to tackle this issue is to incorporate the object scale information to condition the generation, which we leave for future work.

\section{Conclusion}
In this work, we introduce a simple yet highly effective approach for handling few-shot setting in segmentation tasks using an inpainting diffusion model. While we only conduct experiments for few-shot segmentation for remote sensing dataset, it is straightforward to adopt our approach for other task, such as object detection, as well as other domains, such as medical imaging and autonomous driving, where annotated data is scarce.

\newpage
\bibliography{iclr2025_conference}
\bibliographystyle{iclr2025_conference}

\newpage
\appendix
\section{Generation Samples}
\label{appendix:samples}
\begin{figure}[ht]
    \centering
    \includegraphics[width=\textwidth]{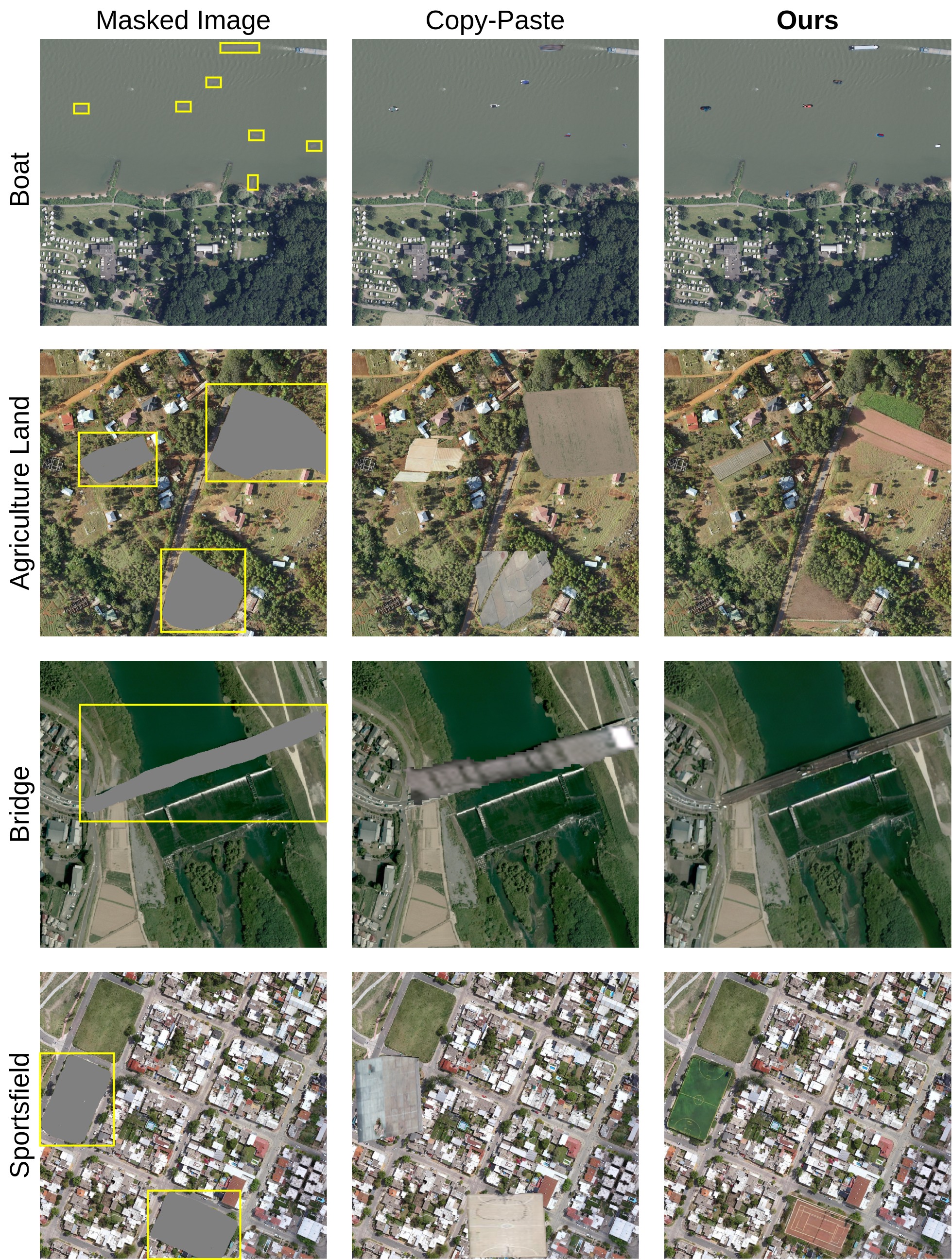}
    \caption{Comparison of inpainting methods for remote sensing: masked image (left), copy-paste (middle), and our method (right), showcasing realistic painted images for boats, agricultural land, bridges, and sportsfields.}
    \label{fig:samples}
\end{figure}

\section{Additional Experiment Details}

\subsection{Dataset}
\label{appendix:dataset_details}
The original few-shot set of the OpenEarthMap dataset comprises 408 samples, which are split into 258, 50, and 100 samples for training, validation, and test set, respectively. There are 7 base classes in the training set, 4 novel classes in the validation set, and another 4 novel classes in the test set. For our experiments, we select four of the eight novel classes to represent objects with varying complexity and scale. To generate the new samples, for each class, we randomly select 10 images from the training set to act as $\mathbf{I}_b$ and use the support set as $\mathbf{I}_r$. From these 10 images, we generate approximately 1000 unique variations, which are then used to train the segmentation models.

\subsection{Training Details}
\label{appendix:training_details}
During the fine-tuning of Stable Diffusion on the SAMRS dataset, the parameters of the image encoder used for conditioning are kept frozen, while the MLP and Stable Diffusion model remain trainable.

In the experiments, we use the following configurations for the baseline models:
\begin{itemize}
    \item YOLOv11, X variant, pretrained on COCO dataset \citep{lin2014microsoft}
    \item SegFormer, B5 variant, pretrained on Cityscapes dataset \citep{cordts2016cityscapes}
    \item Mask2Former, Large variant, pretrained on Cityscapes dataset \citep{cordts2016cityscapes}
\end{itemize}
Using the corresponding training strategies, we train each model for 40 epochs, batch size of 32, and learning rate of 5e-5. We also use augmentations such as random cropping, random horizontal and vertical flipping, random rotation, random brightness scaling, and random gaussian blur.

\section{Combining Variations}
\label{appendix:combine}
Given the generation results for $N$ different regions $\{\mathbf{C}_p^n\}_{n=1}^N$, where each region has $L$ variations, we aim to generate all possible combinations of these regions. Each combination corresponds to a different variation of the final image. The idea is to combine selected regions from all possible subsets of the $N$ regions, while considering all possible variations within each region.

The formalized steps are as follows:
\begin{enumerate}
    
    \item \textbf{Binary Mask Representation}: 
    To form the combinations, we generate all possible binary masks of length $N$, where the $n$-th bit corresponds to $\mathbf{C}_p^n$. A bit value of 1 indicates that the corresponding region is included, and a bit value of 0 indicates that the region is excluded. 
    
    \item \textbf{Combining Regions}: 
    For each binary mask, we generate a variation by:
    \begin{itemize}
        \item Including only those regions $\mathbf{C}_p^n$ for which the corresponding bit in the mask is 1.
        \item For each included region, we select one of its $L$ variations. Therefore, the number of combinations for a specific binary mask is $L^k$, where $k$ is the number of selected regions, \textit{i.e.}, the number of 1's in the binary mask.
    \end{itemize}

\end{enumerate}

The total number of unique variations can be computed by summing the possible combinations for all subsets of regions. Specifically, for any subset of size $k$, there are $\binom{N}{k}$ ways to select $k$ regions, and for each of these subsets, there are $L^k$ possible variations. Therefore, the total number of unique variations is $\sum_{k=1}^{N} \binom{N}{k} L^k$.

\end{document}